# A versatile platform for graphene nanoribbon synthesis, electronic decoupling, and spin polarized measurements

Aleš Cahlík[1]*, Danyang Liu[1], Berk Zengin[1], Mert Taskin[1], Johannes Schwenk[2], Fabian Donat Natterer[1]*

[1]Department of Physics, University of Zurich, Winterthurerstrasse 190, CH-8057 Zurich
[2]Institute of Physics, EPFL, Station 3, CH-1015 Lausanne

*Corresponding authors: ales.cahlik@physik.uzh.ch, fabian.natterer@physik.uzh.ch

**Abstract**
The on-surface synthesis of nano-graphenes has led the charge in prototyping structures with perspectives beyond silicon-based technology. Following reports of open-shell systems in graphene-nanoribbons (GNR), a flurry of research activities is directed at investigating their magnetic properties with a keen eye for spintronic applications. Although the synthesis of nano-graphenes is usually straightforward on gold, it is difficult to use it for electronic decoupling and spin-polarized measurements. Using a binary alloy $Cu_3Au(111)$, we show how to combine the efficient gold-like nano-graphene formation with spin polarization and electronic decoupling known from copper. We prepare copper oxide layers, demonstrate thermally and tip-assisted synthesis of GNR and grow thermally stable magnetic Co islands. We functionalize the tip of a scanning tunneling microscope with carbon-monoxide, nickelocene, or attach Co clusters for high-resolution imaging, magnetic sensing, or spin-polarized measurements. This versatile platform will be a valuable tool in the advanced study of magnetic nano-graphenes.

**Introduction**
Through experience, we have come to accept that one cannot always have one's cake and eat it, too. Unfortunately, this is also true in scientific endeavors where it even may be self-induced. This partially originates in the way investigations grow from the bottom-up, expanding on previous progress and following a curiosity driven path. Such style of research works well until a top-down challenge arises that may be incompatible within the so far used framework. We see parallels to this conundrum in research following the pioneering works on on-surface synthesis of graphene nanoribbon (GNR) (*1*) and chemically tailored nano-graphenes (*2*). For their synthesis, the overwhelming number of contributions have converged on the coinage metals gold (*3–15*) or copper (*2, 16–19*). Although both provide simple preparation, gold is deemed favorable for tailored synthesis because it reliably facilitates polymerization directly at the halogenated carbon site. For example, on-surface reaction of 10,10′-dibromo-9,9′-bianthracene (DBBA) results in the growth of straight $N = 7$ armchair graphene nanoribbons (7-AGNR) on Au(111) (*1*) whilst on Cu(111) partially chiral nanoribbons are formed (*17, 20*). However, with the emergence of magnetic signatures in carbon-based systems (*6–8*, *12*, *13*, *21*, *22*), further investigation may depend on our ability to provide orbital-imaging, spin-polarization and decoupling from itinerant electrons that are hard to simultaneously satisfy for either substrate. When looking for model-systems of spin-polarization, we notice in cobalt nanoislands on copper (*23*, *24*) a similar monoculture with a heavy focus on this platform despite being plagued by the rapid intermixing of Co and Cu at room temperature (*25*), which precludes its use for thermally induced nano-graphene formation. Perhaps more liberty and flexibility are found in investigations of decoupling layers such as $Cu_2N/Cu$ (*26, 27*), MgO/Ag (*28, 29*), or NaCl/Cu (*30*) because they appear motivated by top-down questions about the pristine properties of single atoms or molecules. A natural question is therefore to ask whether tailored on-surface synthesis, spin-polarization, or decoupling are exclusive and if not whether we can identify systems that simultaneously host the properties that made one specific platform so popular.

Here we establish with $Cu_3Au$ such a system that combines the key factors of the coinage metals Au and Cu. We show the preparation of clean $Cu_3Au(111)$ surfaces and the growth of copper oxide decoupling layers. Using the prototypical halocarbon precursor 10,10′-dibromo-9,9′-bianthracene (DBBA), we demonstrate their polymerization and cyclodehydrogenation into $N = 7$ armchair GNRs. We show that the latter step can either be temperature induced or via deliberate tip-manipulation using the scanning tunneling microscope (STM). We use high-resolution STM with carbon-monoxide (CO) functionalized tips to verify the successful synthesis into 7-AGNR and observe a spontaneous formation of GNRs exhibiting zero-bias peaks, suggestive of a Kondo resonance. In order to demonstrate spin-polarization on $Cu_3Au$, we prepare cobalt nanoislands that are thermally and magnetically stable and serve as an easy means to produce spin-polarized tips. Together, these salient features demonstrated in our work equip the community with a versatile platform to tackle pressing top-down questions in surface science and in the advanced investigation of carbon-based systems.

## Results
### $Cu_3Au$ substrate and its oxidation
We prepare clean surfaces of $Cu_3Au$ following previous work (*31*), using standard surface cleaning procedures described in the Methods Section. Figure 1A shows an STM overview of a $Cu_3Au(111)$ surface termination with large terraces routinely exceeding 150 nm width and separated by monoatomic steps. High-resolution imaging using a carbon-monoxide functionalized tip shows the atomic lattice and the (2x2) supercell of the $L1_2$ ordered phase (Figure 1B-C)(*32*). Similar to the pure coinage metals, $Cu_3Au(111)$ exhibits a nearly-free electron like surface state of effective mass $m^*/m_e = 0.31 \pm 0.02$ and band onset of $E_0=0.42$ eV below the Fermi level when measured using quasiparticle interference imaging (*33*), comparable to the observed value in previous work using angle-resolved photoelectron emission spectroscopy (*34*). The darker line-features in Figure 1A and E occasionally form complete hexagonal networks with a long-range periodicity of about $(36 \pm 3)$ nm (Figure S1). We propose that these networks are associated with the previously reported 29 nm surface reconstruction, which had been noticed as faint spots in low-energy electron diffraction (LEED) experiments and described as a "herringbone-like" reconstruction (*31*) (for details see Supplementary material).

Having established the preparation of clean surfaces of $Cu_3Au$ and verified their properties, we proceed to demonstrate the growth of copper oxide overlayers. We deliberately dose small amounts of molecular oxygen while annealing the crystal (see Methods section) to form sub-monolayer patches of an oxidized substrate (Figure 1E). The threefold symmetric contrast (Figure 1F, G) is reminiscent of cuprous oxide in previous reports for $Cu_2O/Pt(111)$ (*35*). Furthermore, the spectroscopic signatures (Figure 1H) of the oxide patch show a characteristic bandgap of about 1.3 eV in good agreement with observations for $Cu_2O/Au(111)$ (*36*). The oxide formation is further confirmed by a distinct change in LEED pattern after the oxygen exposure (Figure 1I). We also observe a formation of oxide layers from interstitial oxygen that naturally segregates towards the surface during repeated and prolonged annealing at temperatures above 600 °C. In this case, two coexisting phases are formed (Figure S2). Besides the threefold symmetric phase observed after the deliberate exposition, a new stripe-like phase emerges exhibiting a similar bandgap in tunneling spectroscopy. We have found no direct precedent for this phase, but $Cu_2O$ is known to constitute a wealth of structures including phases of reduced symmetry, such as in the Pt(111) system (*37*).

**Synthesis of nano-graphenes**
We proceed to verify the utility of $Cu_3Au(111)$ for the on-surface synthesis of nano-graphenes. The most prominent example is the growth of graphene-nanoribbons (GNR) from DBBA precursor molecules on different coinage metal substrates (*1*, *17*, *20*). We establish the growth of straight $N = 7$ armchair GNRs on $Cu_3Au(111)$ by pursuing the common two-step formation procedure (Figure 2A). To that end, we deposit the precursor DBBA molecule at the substrate kept at 200 °C, leading to the dehalogenation and subsequent polymerization of the bi-anthracene moieties into one-dimensional chains. The second annealing step to a higher temperature of 300 °C yields through cyclodehydrogenation the fully formed GNR (Figure 2B). To obtain a peek into the two-step synthesis, we interrupt the high temperature annealing after 15 minutes to ensure the coexistence of fully formed GNRs and the polymerized but still hydrogenated chains. The polyanthrylene chains can be distinguished from the GNR by the bead-like protrusions and larger apparent-height as reported in previous works (*1*). We verify the structure of the 7-AGNR using high-resolution STM imaging with a CO functionalized tip (Figure 2C). The image clearly reveals the straight 7-AGNR structure corresponding to the nanoribbon formation on Au(111) (*1*) in contrast to the growth of partially chiral nanoribbons from DBBA on Cu(111) (*17*).

In addition to pristine 7-AGNRs, we also observe the spontaneous growth of defective ribbons (Figure 2D). The interest in GNR defects lies in their connection to single electron spins in open-shell structures exhibiting magnetic signatures (*4*). Similarly, we observe the characteristic zero bias peaks in scanning tunneling spectroscopy for several defective structures, suggestive of Kondo resonances (inset Figure 2D) - an established feature of the stabilized radical.

Remarkably, when scanning at elevated bias, we notice instabilities in the protrusions of polyanthrylene chains, indicating a mechanism for their deliberate dehydrogenation. We accordingly use the tip of the STM for a targeted cyclodehydrogenation by stripping-off hydrogen from selected parts of the polymerized chains (Figure 2E). The thusly formed GNRs have the same apparent height as the ones formed by the second annealing step.

**Magnetic properties of Cobalt nanostructures and spin-polarization**
We finally demonstrate the sub-monolayer (ML) growth of cobalt nanoislands on $Cu_3Au(111)$ and characterize their magnetic signatures. In contrast to the fast intermixing that is typical of Co/Cu(111) already at room-temperature, the Co/$Cu_3Au$ interface is thermally robust. We do not observe signs of intermixing for substrate temperatures up to about 300 °C. After deposition of 0.3 ML Co at about 200 °C, we observe the growth of round and triangular Co islands of ~3.4 Å apparent height (Figure 3A). Compared to the cobalt islands on Cu(111), the islands appear smaller with the edge length spanning from 20 nm to a few nm as a result of a larger lattice mismatch (5% for Co-Cu$_3$Au, 2% for Co-Cu) (*38*).

For the magnetic characterization, we focus on triangular islands due to their resemblance with the model system Co/Cu(111)(*23*). To that end, we use nickelocene (NiCp$_2$) molecules, whose spin-excitation from the $S = 0$ ground to the $S = \pm1$ excited state is sensitive to the magnetic or exchange field applied along the molecular axis (*39*, *40*). At zero-field, both $S = \pm1$ transitions are degenerate but in an applied field, the Zeeman splitting separates the single step in the tunneling conductance into two. Exposing a NiCp$_2$ functionalized tip to the exchange field of a Co island on Cu$_3$Au results in clearly observable splitting of the step feature in STS when we move the NiCp$_2$ tip closer to the Co island (Figure 3B, C). The proximity leads to an enhanced exchange field and concomitantly to a larger Zeeman splitting. This exchange interaction related shift in the spin-excitation step is equivalent to previous reports for NiCp$_2$ measured against Co islands on Cu(111) (*39*).

To establish the magnetic bistability of our Co islands, we select two triangular islands of identical stacking and similar size to avoid structural ambiguities (Figure 3D). To further ensure their electronic equivalency without strain induced effects (*41*, *42*), we measure the tunneling spectroscopy with a spin-averaging tip. The two selected islands are spectroscopically equivalent and share the same *d*-level position at -290 mV of comparable intensity (Figure 3F, G). We note that the average *d*-level position at -310 mV in tunneling spectroscopy (see the histogram for tunneling spectra taken above 66 random triangular islands in Figure 3E) is comparable with the reported peak position for Co on Cu(111) (*42*), although the shift towards lower energies could be expected due to larger lattice mismatch between cobalt and Cu$_3$Au. We tentatively attribute this to edge induced effects on the peak position due to the smaller island size.

Using a spin-polarized tip on the same two islands, the intensity of the *d*-level peak becomes distinct between the orange and red island (Figure 3H, I), showing an unequivocal signature of tunnel magnetoresistance stemming from the different magnetic orientations of the two islands with respect to the magnetization of the tip. This is further supported by a contrast reversal between the two islands after a spontaneous switch of the tip magnetization (Figure 3J, K).

For completeness, we finally demonstrate how to readily produce spin-polarized tips by simply picking up a Co island from the Cu$_3$Au substrate which transfers Co to the tip apex (see Methods for detail). This procedure leaves the neighboring structures intact (Figure 3L), causing minimal disruption to local details.

**Discussion**

When we look at the overall properties of Cu$_3$Au, we identify interesting opportunities and perspectives. The most notable behavior is the thermal stability of the interface to increased temperature, which is a crucial asset for temperature driven on-surface synthesis. This holds, even in presence of the copper oxide overlayer and the magnetic Co islands; the latter being thermally stable to 300 °C. The value of this stability is immediately clear when we compare them to the requirements for dehalogenation and cyclohydrogenation as elementary steps in Ullmann coupled systems (Figure 2A). Our work thus combines the controlled on-surface synthesis of GNR known from Au(111) with the magnetism typical of Co/Cu(111). The absence of the short-range herringbone reconstruction on Cu$_3$Au(111) is helpful in reciprocal space analyses of synthesis steps and the possibility for magnetic Co islands offers facile spin-polarization and magnetic references for the study of open-shell GNR. In addition, the copper oxide layer can serve as a decoupling layer that isolates the GNRs magnetic signature from their interaction with itinerant electrons. The $Cu_3Au$ system can also stabilize a lattice matched copper-nitride for which electronic decoupling was successfully demonstrated previously with single atoms (*43*, *44*).

The tip functionalization on $Cu_3Au$ is straightforward which is valuable for the verification of synthesis steps or to enhance certain contrast modes. We demonstrate this with the routine use of CO, $NiCp_2$, and Co cluster functionalized tips for high-resolution imaging, magnetic imaging, and spin-polarization, respectively. The facile tip-functionalization promotes further work on magnetism relying on direct tunneling (*45*) or STM enabled electron-spin resonance experiments (*29*). In addition to functionalization, we also show how the tip can be used to steer the second GNR synthesis step by deliberately removing selected hydrogen bonds of polymerized molecules. This invites further work utilizing defects or barriers in circular or one-dimensional polymerized carbon structures.

## Materials and Methods
### Sample Preparation
The $Cu_3Au(111)$ crystal was purchased from Surface preparation laboratory (SPL). We mount it on a ferromagnetic sample plate to allow for the attachment of permanent magnets for field-dependent studies (*46*) and prepare atomically clean surfaces analogous to literature (*31*) by repeated cycles of $Ar^+$ ion bombardment and annealing (3 keV, 0.8 µA/cm$^{-2}$, 550 °C). To verify the impact of the order/disorder transition, we perform extended annealing procedures (16 hours at 340 °C) to promote ordering of the $L1_2$ structure (*32*). We find no evidence that the ordering level influences the GNR formation or the cobalt island growth. The oxidized surface was prepared by annealing a clean $Cu_3Au$ substrate at 400 °C while exposing to $8\times10^{-7}$ mbar $O_2$ atmosphere for 20 minutes. We grow cobalt islands from thoroughly degassed Co rods using a commercial e-beam evaporator (Focus EFM3) at sample temperatures from 60 °C to 300 °C. Increasing the substrate temperature beyond 300 °C results in visible intermixing at the step edges, emergence of vacancies and sinking of the islands into the substrate. For the GNR growth, we dose commercially available 10,10′-dibromo-9,9′-bianthracene precursors (Chemie Brunschwig AG, CAS: 121848-75-7) using an effusion cell (Kentax, TCEorg-3BSC) at a source temperature of 160 °C.

### Tip functionalization
For high-resolution imaging, we functionalize the tip with a carbon monoxide molecule that we pick off the substrate either spontaneously during scanning (Figure 2B) or by stabilizing the tip above the molecule (100 mV, 10 pA), switching off the feedback loop and increasing the bias to 3 V. We dose nickelocene ($NiCp_2$) from a commercially available powder (Chemie Brunschwig AG, CAS: 1271-28-9) directly onto the cooled sample in the microscope head. The nickelocene tip functionalization is performed with feedback loop off, by manually approaching the tip to the molecule at 20 mV bias until a sharp jump in the current channel is observed. For spin-polarized measurements, we deliberately transfer a Co island from the $Cu_3Au$ substrate to the tip. This is done by approaching the island from 20 pA, 50 mV setpoint by 400-500 pm while applying 2-2.5 V bias.

### STM Measurements and Tunneling Spectroscopy
All experiments are performed in ultra-high vacuum using a commercial STM (CreaTec Fischer & Co. GmbH) operating at about 4 K. Our tip is made from a mechanically cut PtIr wire and sharpened by gently plunging it into the Cu$_3$Au sample until we notice sharp steps edges in topographic scans. For point-spectroscopy, we use a conventional lock-in technique at frequency of 932 Hz. For the surface state mapping, we use a multifrequency lock-in amplifier (Intermodulation Products SA, MLA-3) as described in Ref.(*33*).

**Acknowledgments**
We thank T. Diulus, M. Hengsberger, J. Osterwalder, R. Wiesendanger, and R. Fasel for fruitful discussion.

**Funding:**
Swiss National Science Foundation PP00P2_176866 (FDN, DL, BZ, AC)
Swiss National Science Foundation 200021_200639 (FDN, AC)

Swiss Government Excellence Fellowship (AC)
Office of Naval Research N00014-20-1-2352 (FDN, BZ)
UZH Forschungskredit FK-20-093 (DL)

**Author contributions:**
Conceptualization: AC, FDN
Methodology: AC, FDN, JS
Formal analysis: AC, DL, FDN
Investigation: AC, BZ, DL, FDN, MT
Visualization: AC, DL
Data Curation: AC
Writing—original draft: FDN
Writing—review & editing: AC, FDN
Supervision: FDN, AC
Project Administration: FDN
Funding acquisition: FDN, AC, DL

**Competing interests:** The authors declare that they have no competing interests.

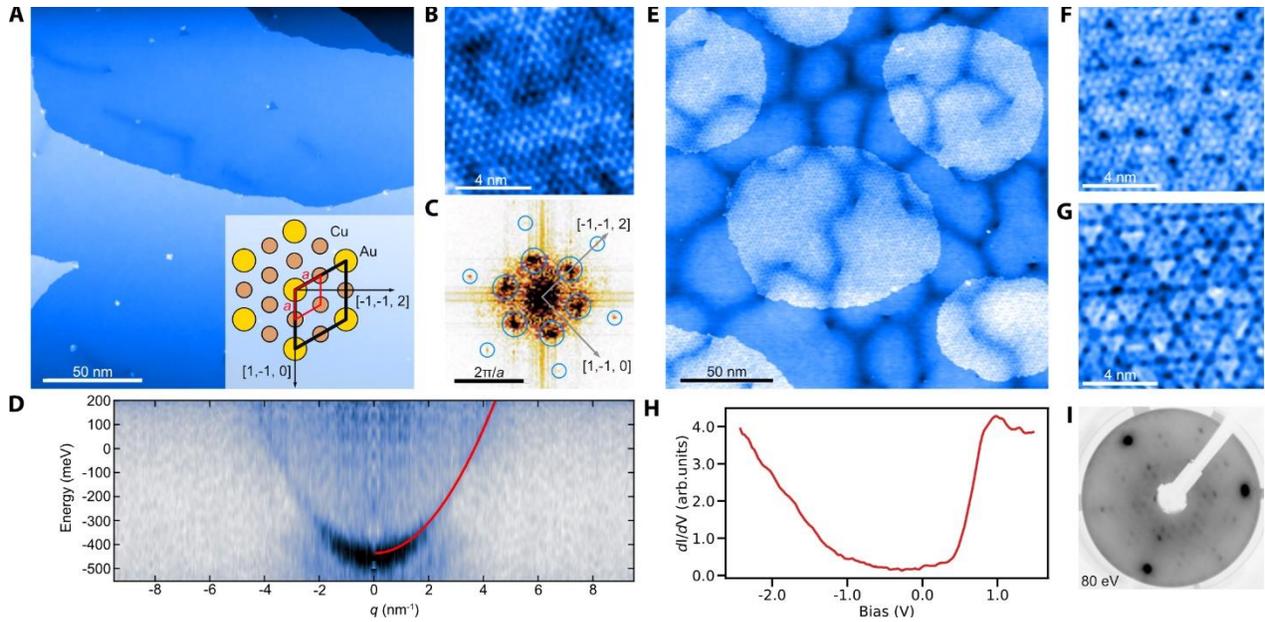

**Figure 1 | Properties of Cu$_3$Au(111) and Copper Oxide Overlayer.** **(A)** Overview topography of clean Cu$_3$Au(111) terraces ($V = 1$ V, $I=140$ pA) with the ideal L1$_2$ surface termination indicated in the inset, partially visible by the atomically resolved image in **(B)** and emphasized in its Fourier transform in **(C)**, showing the atomic lattice (small circles) and the (2×2) supercell (large circles). **(D)** The surface hosts a nearly-free electron like surface state with an effective mass of m*/m$_e$ = 0.31±0.02 as determined by the parabolic fit (red-line) to the dispersion. The dispersion plot was produced from the measurement of the energy dependent local density of states on a field-of-view of 105 nm (drive frequency 1600 Hz, modulation amplitude 0.5 V, offset -0.25 V) **(E)** Topographic image of Cu$_3$Au surface with patches of oxide phase after exposure to molecular oxygen. (-0.5 V, 50 pA) **(F), (G)** Threefold symmetric contrast in zoomed images of the oxide patch for two different bias values. ($V = -0.3$ V and $V = 0.5$ V respectively) **(H)** Point tunneling spectroscopy of the oxide phase showing a gap-size of about 1.3 V. **(I)** LEED image of the oxidized substrate at 80 V beam energy showing spots in addition to the Cu$_3$Au lattice.

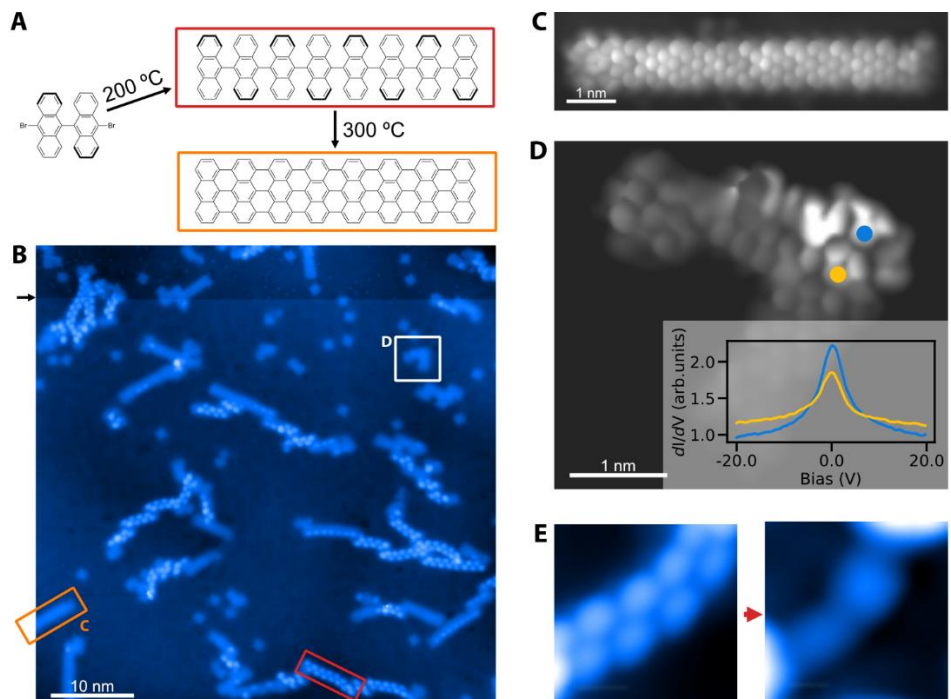

**Figure 2/ On-surface Synthesis of Nano-Graphenes on Cu$_3$Au(111). (A)** Two step 7-AGNR annealing scheme from DBBA precursor molecules. (**B**) Overview topography after dosing DBBA onto the substrate kept at 200 °C and post annealing to 300 °C ($V$ = 50 mV, $I$ = 30 pA). While DBBA readily polymerizes according to the scheme in **(A)**, the cyclodehydrogenation was interrupted after 15 minutes of annealing at the higher temperature to leave some chains polymerized but still hydrogenated (marked by the red rectangle). The black arrow marks a spontaneous tip functionalization with a CO molecule. **(C)** High-resolution STM image of a fully formed $N$ = 7 armchair GNR using a carbon-monoxide functionalized tip. **(D)** Spontaneously formed defective GNR showing a zero-bias peak in tunneling spectroscopy (inset), suggestive of a Kondo resonance. **(E)** The tip of the STM can be used for targeted and deliberate cyclodehydrogenation as shown by the topographies before and after applying current pulses on a polymerized chain ($V$ = 3.5 V, $I$ = 20 nA).

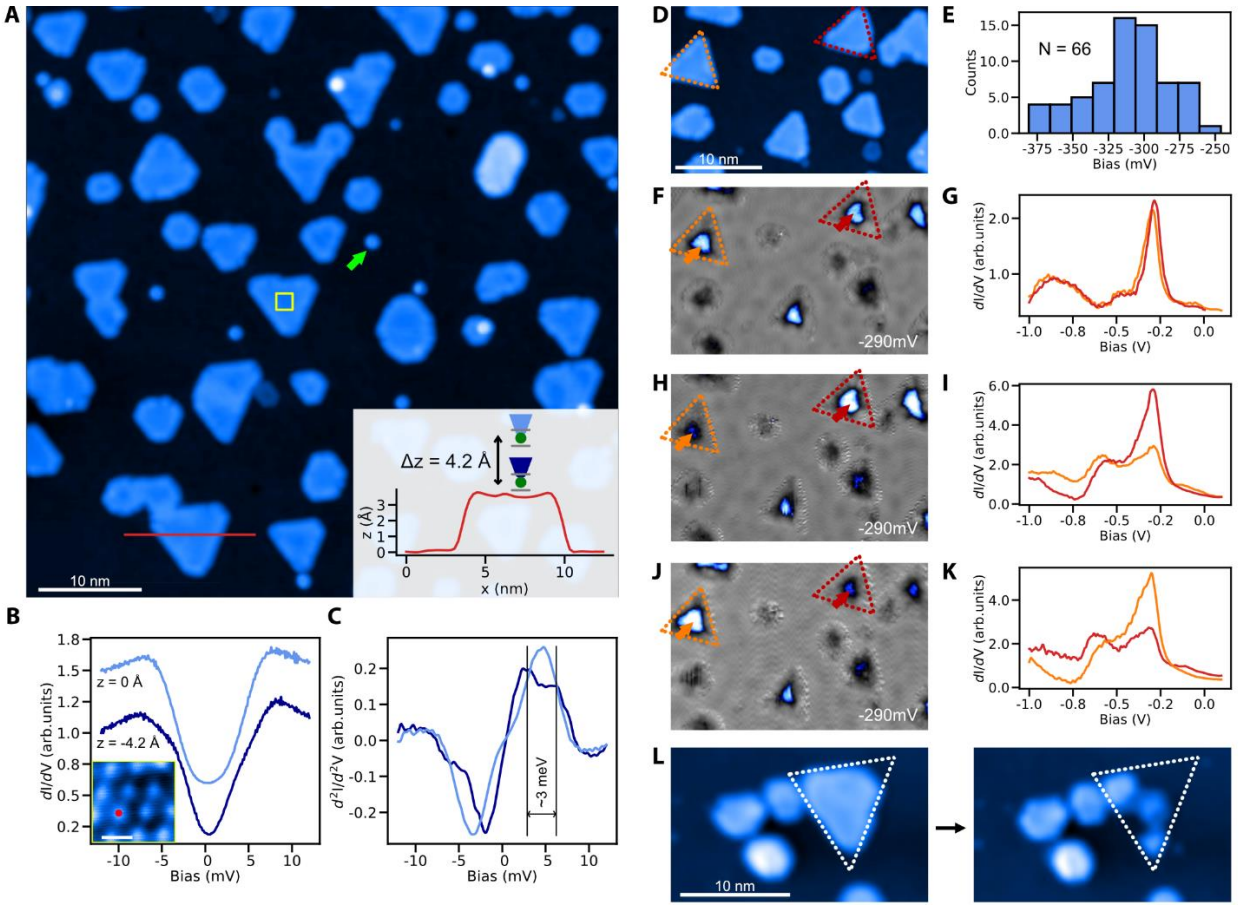

**Figure 3 | Growth of cobalt islands on Cu$_3$Au and facile preparation of spin-polarized tips.**
(**A**) Overview topography of two-monolayer high Co islands that were grown at 200 °C with the line profile across the island shown in the inset ($V$ = 50 mV, $I$ = 5 pA). The circular and triangular islands have different stacking with respect to the substrate. (**B**) Tunneling conductance spectra and (**C**) its derivative taken with nickelocene functionalized tip for two different heights ($\Delta z$ = 4.2 Å, inset of **A**) above a Co atom in the middle of an island (marked in the inset of **B**). The spectrum taken closer to the Co atom shows a splitting of ~3 meV, corresponding to the exchange field in the order of 13 T (39). (**D**) Topography image containing two similarly sized islands (dotted lines) with identical stacking used for the magnetic characterization of their magnetic bistability and spin-polarization. (**E**) Histogram of the *d*-level peak position with a mean of -0.31 V as determined from 66 randomly picked triangular islands. (**F**) Closed loop conductance image taken at -290 mV, measured with spin-averaging tip and corresponding point spectra (**G**) on the orange and red island showing no contrast difference (10 mV modulation). (**H**) and (**I**) Conductance map and tunneling spectra measured over the same region with a spin-polarized tip (10 mV modulation). The intensity of the *d*-level shows a clear difference that is attributed to two distinct magnetic orientations between the two islands and that is also visible as contrast variation in the conductance map. (**J**) and (**K**) Conductance map and tunneling spectra with spin-polarized tip whose magnetization spontaneously switched, leading to a reversed contrast between the two magnetically differently oriented islands. (**L**) Spin-polarized tips can be produced by transferring an island from the substrate to the tip as shown by the before and after topographic images ($V$ = 400 mV, $I$ = 50 pA).